# Electronic Coupling and Charge-Transfer Landscape of Graphene on Ge(001)/Si(001): Multiscale Analysis Assisted by Machine Learning

Pawel Dabrowski*[1], Przemysław Przybysz*[1,2], Maciej Rogala[1], Iaroslav Lutsyk[1], Paweł Krukowski[1], Witold Kozłowski[1], Michał Piskorski[1] , Piotr Milczarski[3,4], Iwona Pasternak[5], Jakub Sitek[5], Marek Kopciuszyński[6], Ryszard Zdyb[6], Jagoda Sławińska[2] and Pawel J. Kowalczyk[1]

[1] Faculty of Physics and Applied Informatics, University of Lodz, Pomorska 149/153, 90-236 Łódź, Poland
[2] Zernike Institute for Advanced Materials, University of Groningen, Nijenborgh 3, 9747 AG Groningen, The Netherlands
[3] Faculty of Technical Physics, Information Technology and Applied Mathematics, Lodz University of Technology, Wólczańska 215, 93-005 Łódź, Poland
[4] Faculty of Mathematics and Computer Science, University of Lodz, Banacha 22, 90-238 Łódź, Poland
[5] Faculty of Physics, Warsaw University of Technology, Koszykowa 75, 00-662 Warsaw, Poland
[6] Institute of Physics, Maria Curie-Sklodowska University, Pl. M. Curie-Sklodowskiej 1, 20-031 Lublin, Poland
*Corresponding Authors*: pawel.dabrowski@uni.lodz.pl and przemyslaw.przybysz@edu.uni.lodz.pl

## Abstract

Understanding and controlling charge transfer at graphene–semiconductor interfaces is essential for the integration of two-dimensional materials into silicon-compatible technologies. Here, we combine ultraviolet photoelectron spectroscopy (UPS), Kelvin probe force microscopy (KPFM), angle-resolved photoemission spectroscopy (ARPES), scanning tunneling spectroscopy (STS) and density functional theory (DFT) to resolve work-function modulation and electronic coupling in graphene grown on Ge(001)/Si(001). UPS and KPFM reveal a spatially non-uniform work-function landscape correlated with the nanofaceted morphology of the substrate. ARPES and DFT calculations for the pristine interface consistently indicate n-type doping and electron transfer from Ge to graphene. In contrast, modeling of the oxidized interface predicts a reversal to p-type doping, providing a plausible explanation for the different doping polarities reported in the literature. Machine-learning-assisted classification of the STS data resolves distinct local electronic regimes, ranging from nearly free-standing graphene on nanofacet tops to more strongly coupled inter-facet regions and nanoribbon-like regions with distinct local electronic responses. By identifying the mechanisms governing local graphene–substrate interactions and charge transfer, our study provides guidelines for tailoring the synthesis process and minimizing defect formation during delamination. These insights support the production of high-quality graphene layers for electronic devices and for transfer as protective coatings for air-sensitive materials.

**TOC Summary**

A combined experimental and theoretical approach reveals a spatially modulated charge-transfer landscape in graphene/Ge(001)/Si(001). Global and local probes, supported by machine-learning analysis, uncover a continuum of graphene–substrate coupling regimes governing work-function variation and electronic properties.

## 1. Introduction

Graphene is a key material for next-generation electronic, optoelectronic and sensing technologies due to its high carrier mobility and sensitivity to the surrounding environment. [1–6] Its atomically thin structure also makes it an ideal platform for van der Waals (vdW) heterostructures [7–11] and an effective encapsulation layer for protecting air-sensitive two-dimensional materials from oxidation.[12–14] However, realizing these capabilities in scalable technologies requires wafer-scale growth methods compatible with complementary metal–oxide–semiconductor (CMOS) processing.[15–17]

Among available synthesis approaches, chemical vapor deposition (CVD) enables wafer-scale graphene growth.[18–21] However, graphene grown on Cu or Ni foils requires transfer to insulating substrates, introducing contamination and structural defects that degrade device performance.[22,23] Epitaxial growth on SiC eliminates transfer but leads to strong graphene–substrate interactions that alter its electronic structure.[24–28] Direct growth on insulating or dielectric substrates such as $Al_2O_3$, $SiO_2$, MgO and $Si_3N_4$ has also been explored.[29–33] However, their limited catalytic activity toward hydrocarbon decomposition often results in discontinuous or defect-rich graphene films.[30,31,34] In addition, substrate polarity and surface roughness, particularly in the case of $Al_2O_3$, may hinder epitaxial alignment, leading to smaller grain sizes and increased defect densities.[33,34] For these reasons, direct CVD growth on dielectric substrates remains challenging for high-mobility device applications. The limited structural quality of such graphene layers may also reduce their suitability as growth platforms or protective coatings for surface sensitive 2D materials.

In this context, semiconducting germanium (Ge) has emerged as a promising, catalyst-free and CMOS-compatible substrate.[16,35,36] Ge does not form stable carbides and supports direct CVD growth of metal-free graphene. The close match of thermal expansion coefficients between graphene and Ge minimizes wrinkling, yielding continuous, high-quality films.[35] Moreover, graphene grown on Ge is free from metallic contamination, enhancing compatibility with silicon-based device fabrication.[37]

The Ge(001)/Si(001) heterostructure is particularly attractive because epitaxial Ge layers can be grown on large-diameter Si wafers using existing semiconductor infrastructure.[16,17,37,38] During CVD growth, however, the Ge(001) surface undergoes reconstruction and faceting, forming nanoscale domains separated by valleys.[17,39,40] While these features enable continuous graphene coverage, they also introduce lateral variations in graphene–substrate interactions.[41–43] The alternating surface dimer orientation further leads to multiple graphene domain orientations, resulting in nanoscale variations of local electronic properties.[36,41,43,44]

Although the structural and transport properties of graphene on Ge are well established, the electronic doping polarity remains controversial. Graphene on Ge(001) has been reported as n-type, nearly intrinsic, or even p-type depending on growth conditions.[16,44] These discrepancies are often attributed to interfacial oxidation, although direct spectroscopic evidence of $GeO_x$ is frequently lacking.[41,45] As a result, the nanoscale mechanisms governing charge transfer and work-function modulation in graphene/Ge(001) systems remain poorly understood, particularly at the nanoscale.

Here, we present a comprehensive multiscale investigation of the graphene/Ge(001) interface, combining photoemission spectroscopy, scanning probe microscopy, density functional theory (DFT) and machine-learning-assisted analysis. Ultraviolet photoemission spectroscopy (UPS) and Kelvin probe force microscopy (KPFM) characterize the global and local work-function landscape, while angle-resolved photoemission spectroscopy (ARPES) and DFT confirm electron transfer from Ge to graphene. At the nanoscale, scanning tunneling spectroscopy (STS) combined with machine learning identifies distinct coupling regimes across the faceted surface, linking morphology with local electronic properties.

These results provide a unified picture of how surface reconstruction, charge transfer and environmental effects govern the work-function and doping landscape of graphene on Ge(001)/Si(001). By bridging global and local measurements, this study resolves previous discrepancies in doping polarity and establishes a general framework for analyzing charge redistribution at graphene–semiconductor interfaces. Detailed insight into the graphene–germanium interface is essential for optimizing the growth of high-quality graphene layers suitable for CMOS-compatible integration and for subsequent transfer as protective coatings onto air-sensitive materials, including topological insulators.[11] By identifying the mechanisms governing local graphene–substrate interactions and charge transfer, our study provides guidelines for tailoring the synthesis process and minimizing defect formation during delamination. These findings not only advance the fundamental physics of graphene–semiconductor interfaces but also support the development of scalable graphene-based platforms for electronic and optoelectronic applications.

## 2. Results and discussion

### 2.1. Work Function Landscape and Local Electrostatic Potential of the Graphene/Ge(001) Interface

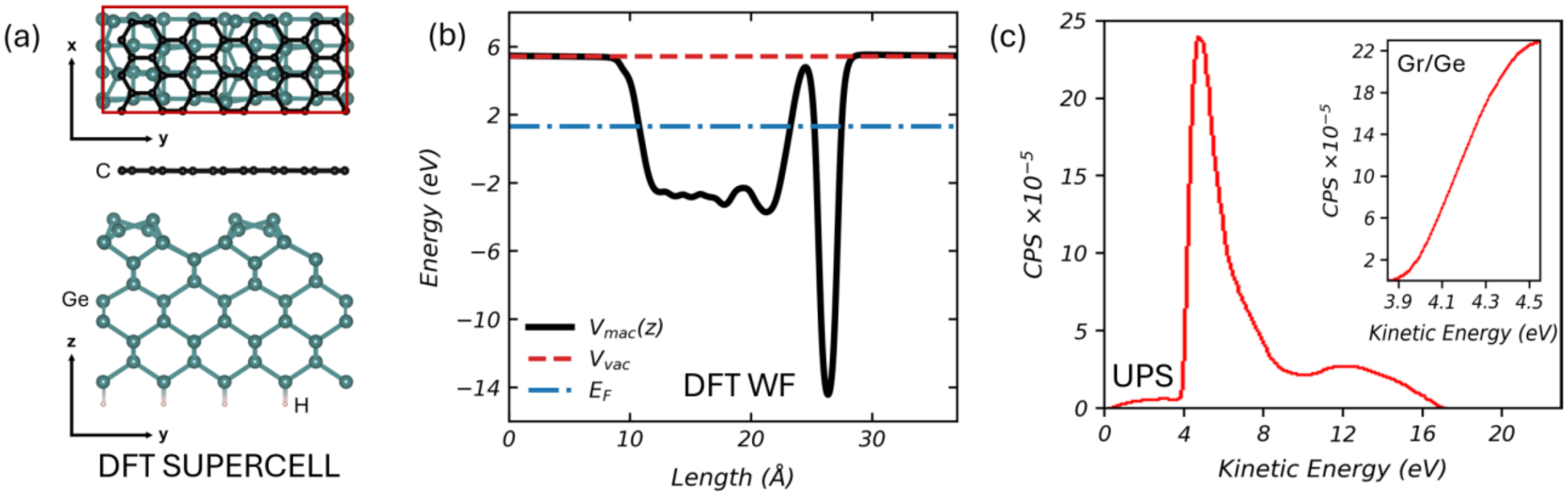

***Figure 1. Structural model and work function determination of graphene/Ge(001)****. (a) Atomic model of the graphene/Ge(001) supercell (128 atoms) used for DFT calculations. (b) Planar-averaged electrostatic potential and resulting work function (WF) determination, yielding a theoretical value of 4.11 eV. (c) UPS spectra showing the secondary electron cutoff used to determine the experimental WF of the graphene/Ge(001) sample (3.9 eV, red curve).*

We first establish a theoretical reference for the work function of an idealized clean graphene/Ge(001) interface. For this purpose graphene monolayer on Ge(001)–(2 × 1) interface was modeled as a supercell containing 128 atoms (Fig.1a). Previous studies have shown that flattened nanofacets on Ge(001) commonly exhibit (2 × 1) and (1 × 2) reconstructions[16] which are consistent with the surface geometry used in our model. Following DFT geometry optimization, the electrostatic potential was evaluated along the direction perpendicular to the surface. The vacuum level, ($V_{\mathrm{vac}}$), was determined from the plateau of the planar-averaged electrostatic potential in the vacuum region and the work function was calculated as ($\Phi = V_{\mathrm{vac}} - E_F$), following the standard slab approach.[46] The calculated work function of the graphene/Ge(001) interface was 4.11 eV (Fig. 1b). The global WF was experimentally determined using ultraviolet photoelectron spectroscopy (UPS) (see Methods and Supporting Information Table 1 and S1 figure for details).

The real graphene/Ge(001) surface is, however, not atomically flat. Previous AFM, STM, LEED, STS and LC-AFM studies have shown that graphene growth on Ge(001)/Si(001) induces a nanofaceted morphology composed of hill-like facet tops separated by inter-facet valleys. These studies demonstrated that graphene on the facet tops retains a weakly perturbed electronic structure and higher local conductivity, whereas graphene in the valley regions is more strongly affected by the interaction with germanium.[43] In the following, we therefore distinguish between nanofacet tops, facet edges and inter-facet valley regions when discussing local variations in the work function and electronic coupling. However, explicitly modeling high-index surfaces was computationally demanding due to the excessive dimensions of the supercell containing both graphene and germanium. Therefore, our theoretical calculations were limited to the idealized flat interface and are representative primarily of the flat regions of the morphology.

Figure 1c shows a sloped secondary-electron cutoff (SECO) for graphene/Ge(001), which is broadened compared to homogeneous reference materials like gold, indicating a distribution of local work functions (WFs). For laterally heterogeneous surfaces, the SECO does not necessarily represent a simple arithmetic average of the local work functions, because electrostatic patch potentials can modify the trajectories of low-energy secondary electrons. Consequently, contributions from different surface regions can broaden and distort the cutoff.[47–49] Using a linear extrapolation of the descending cutoff edge gives an effective UPS-derived WF of approximately 3.8–3.9 eV. This value is slightly lower than those measured for graphene on SiC and graphite under the same experimental conditions (see Supporting Information, Table 1). However, when accounting for the aforementioned measurement factors and modeling limitations, this result is reasonably consistent with our DFT predictions for the clean graphene/Ge(001) interface.

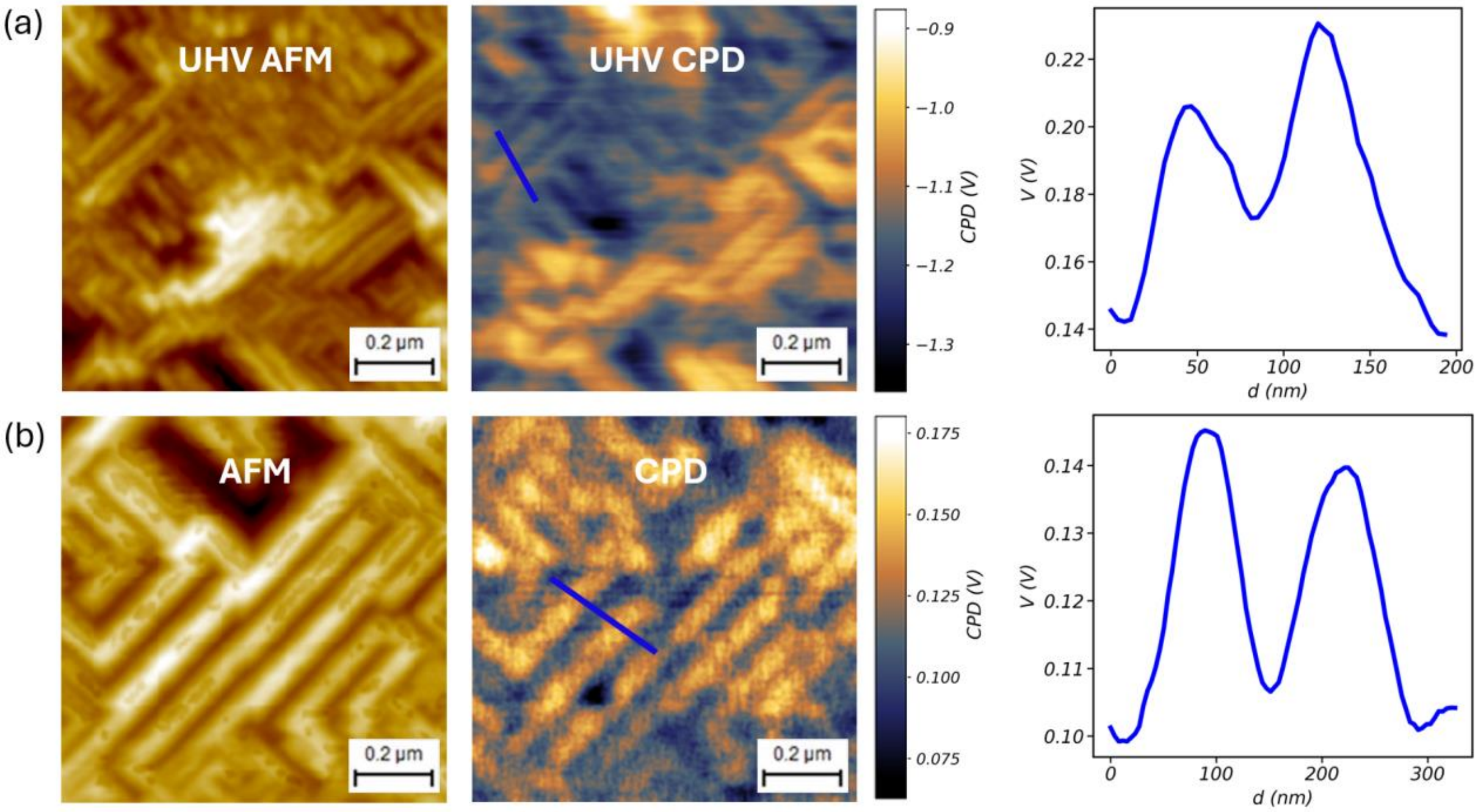

***Figure 2. KPFM measurements of graphene/Ge(001)/Si(001) acquired under*** *(a) ultra-high-vacuum (UHV) and (b) ambient conditions. Each panel shows the AFM topography (left), the corresponding CPD map (middle) and a CPD line profile acquired along the blue line indicated in the map (right). The CPD contrast between nanofacet tops and inter-facet valleys is approximately 35-40 mV under both UHV and ambient conditions. Using the sign convention adopted here, the inter-facet valleys correspond to a higher local work function than the nanofacet tops.*

To resolve these local WF variations, KPFM measurements were performed under UHV and ambient conditions (Figure 2). The topographic images reveal the characteristic nanofaceted morphology of graphene/Ge(001), composed of relatively flat nanofacet tops separated by narrow inter-facet valleys. In the following discussion, these regions are referred to as nanofacet tops, facet edges and inter-facet valleys. In KPFM, the measured contact potential difference (CPD) reflects the difference between the work functions of the tip and the sample. Assuming a constant tip work function, spatial variations in CPD can therefore be used to map relative local WF variations across the surface.

The UHV experiments were conducted in the same chamber as UPS, enabling a direct comparison between the global UPS-derived WF and the local electronic structure contrast prior to atmospheric exposure. In UHV, frequency-modulation KPFM (FM-KPFM) was employed to measure the local CPD with high spatial resolution (Figure 2a). The topography image and the corresponding CPD map show a clear correlation between surface morphology and WF values: nanofacet tops and inter-facet valleys exhibit distinct CPD levels. The CPD line profile acquired across the nanofaceted surface reveals a characteristic contrast of approximately 40 mV between nanofacet tops and inter-facet valleys. The inter-facet valleys correspond to a higher local WF than the nanofacet tops.

After exposure to air, the same sample was remeasured by amplitude-modulation KPFM under ambient conditions (Figure 2b). A comparable morphology-correlated CPD contrast of approximately 35–40 mV was observed between nanofacet tops and inter-facet valleys. Although the absolute CPD values cannot be directly compared between the UHV[38,39] and ambient measurements because different experimental configurations were used, the relative contrast and its correlation with the nanofaceted morphology are preserved. This observation supports the assignment of the local electrostatic variations primarily to the underlying graphene/Ge interface rather than to atmospheric adsorbates alone.

The combined DFT, UPS and KPFM results indicate that the graphene/Ge(001) interface exhibits a spatially heterogeneous electrostatic landscape rather than a uniform work function. The calculated WF of 4.11 eV describes the idealized clean and flat graphene/Ge(001) interface, whereas UPS yields an effective WF of approximately 3.8–3.9 eV for the real nanofaceted surface. KPFM further resolves a morphology-correlated local CPD contrast of approximately 35-40 mV between nanofacet tops and inter-facet valleys.

### 2.2 Charge-Transfer Mechanisms and Electronic Coupling at the Graphene/Ge(001) Interface

As demonstrated in the previous section, graphene on Ge(001) exhibits a spatially non-uniform work-function landscape that correlates with the nanofaceted morphology of the substrate. To understand the electronic origin of these variations, we examined the nature of the graphene–Ge interaction and the associated charge-transfer mechanism.

In general, two limiting scenarios can be considered for graphene–substrate interfaces. In the case of strong interaction, commonly referred to as chemisorption, chemical bonding between graphene and the substrate may substantially perturb or even disrupt the characteristic linear dispersion of graphene.[50,51] By contrast, weak interaction, or physisorption, largely preserves the graphene band structure, although charge transfer may still shift the Dirac point relative to the Fermi level and result in either n-type or p-type doping.[51,52] The magnitude and even the direction of charge transfer depend sensitively on the interfacial electronic structure. Therefore, an accurate description of the graphene/Ge(001) interface requires a combination of experimental spectroscopy and theoretical modeling.

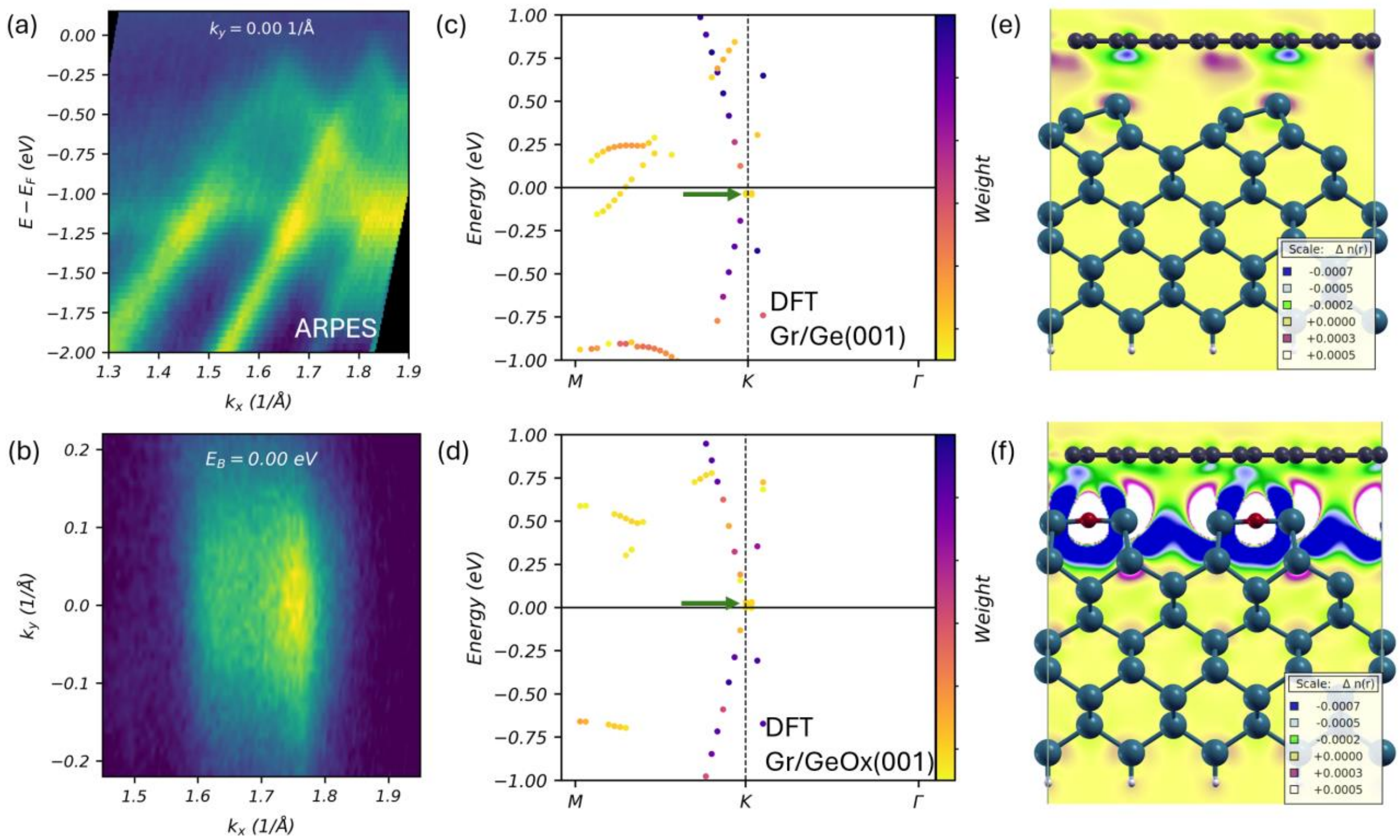

***Figure 3. Electronic structure and charge-transfer of the graphene/Ge(001) interface.*** *(a) ARPES spectrum showing the graphene Dirac point located 0.1 eV below the Fermi level ($E_F$), characteristic of n-type doping. (b) ARPES constant-energy map acquired at $E_F$. (c) DFT-calculated and unfolded band structure of the graphene/Ge supercell, which reproduces the n-type doping with the Dirac point at 0.05 eV below $E_F$. (d) DFT-calculated and unfolded band structure for graphene on oxidized Ge(001), revealing a transition to p-type doping with the Dirac point 0.1 eV above $E_F$. (e) Charge-density-difference map for the pristine interface, illustrating electron accumulation within the graphene layer and depletion at the Ge surface. (f) Charge-density-difference map for the oxidized interface, illustrating electron transfer from graphene toward surface oxygen species. This inversion of charge polarity explains the n-to-p type doping transition induced by oxidation.*

The global electronic properties of the graphene/Ge(001) sample were examined using angle-resolved photoemission spectroscopy (ARPES). Figure 3a shows the characteristic linear dispersion of graphene, indicating that the electronic structure is not dominated by uniformly strong chemisorption. The Dirac point is located approximately 0.1 eV below the Fermi level, revealing n-type doping. This result is consistent with the relatively low effective WF determined by UPS and indicates electron transfer from Ge to graphene. The sharp ARPES features further confirm the high crystalline quality of the graphene layer.

The doping polarity reported for graphene on Ge surfaces varies considerably in the literature, ranging from p-type[16] to n-type [38] or nearly charge-neutral behavior.[16] These discrepancies have been attributed to differences in the growth conditions, surface reconstruction, defect density and the presence of interfacial oxygen bonded to Ge. To clarify the origin of the charge transfer in our samples, we performed DFT calculations for both pristine and oxidized graphene/Ge(001) interfaces.

For the pristine graphene/Ge(001) interface with a (2 × 1) surface reconstruction, the unfolded band structure shown in Figure 3c places the Dirac point approximately 0.05 eV below $E_F$, in agreement with the n-type doping observed by ARPES. The corresponding charge-density-difference map (Figure 3e) reveals electron accumulation within the graphene layer and electron depletion near the Ge surface, confirming charge transfer from Ge to graphene. The redistribution is spatially localized near the surface Ge dimers. These results indicate that the clean graphene/Ge interface occupies an intermediate interaction regime: graphene retains its characteristic linear dispersion, while local graphene–substrate coupling remains sufficiently strong to induce measurable charge redistribution and WF modulation.

Considering the different doping polarities reported for graphene on Ge surfaces, we also examined whether interfacial oxidation[45] could account for this discrepancy. For this purpose, we modeled an oxidized graphene/Ge interface using an established Ge–O bonding configuration.[53,54] The unfolded band structure shown in Figure 3d reveals a reversal of the doping polarity, with the Dirac point located approximately 0.1 eV above $E_F$. The corresponding charge-density-difference map (Figure 3f) shows electron depletion in graphene and accumulation near the surface oxygen atoms, indicating charge transfer from graphene toward the oxidized Ge surface. Oxygen passivates the Ge dangling bonds, modifies the interfacial dipole and reorganizes the charge distribution at the interface. The absence of detectable oxidized Ge in the XPS spectra (Figure S2) further supports the interpretation that the observed n-type doping originates predominantly from the non-oxidized graphene/Ge interface.

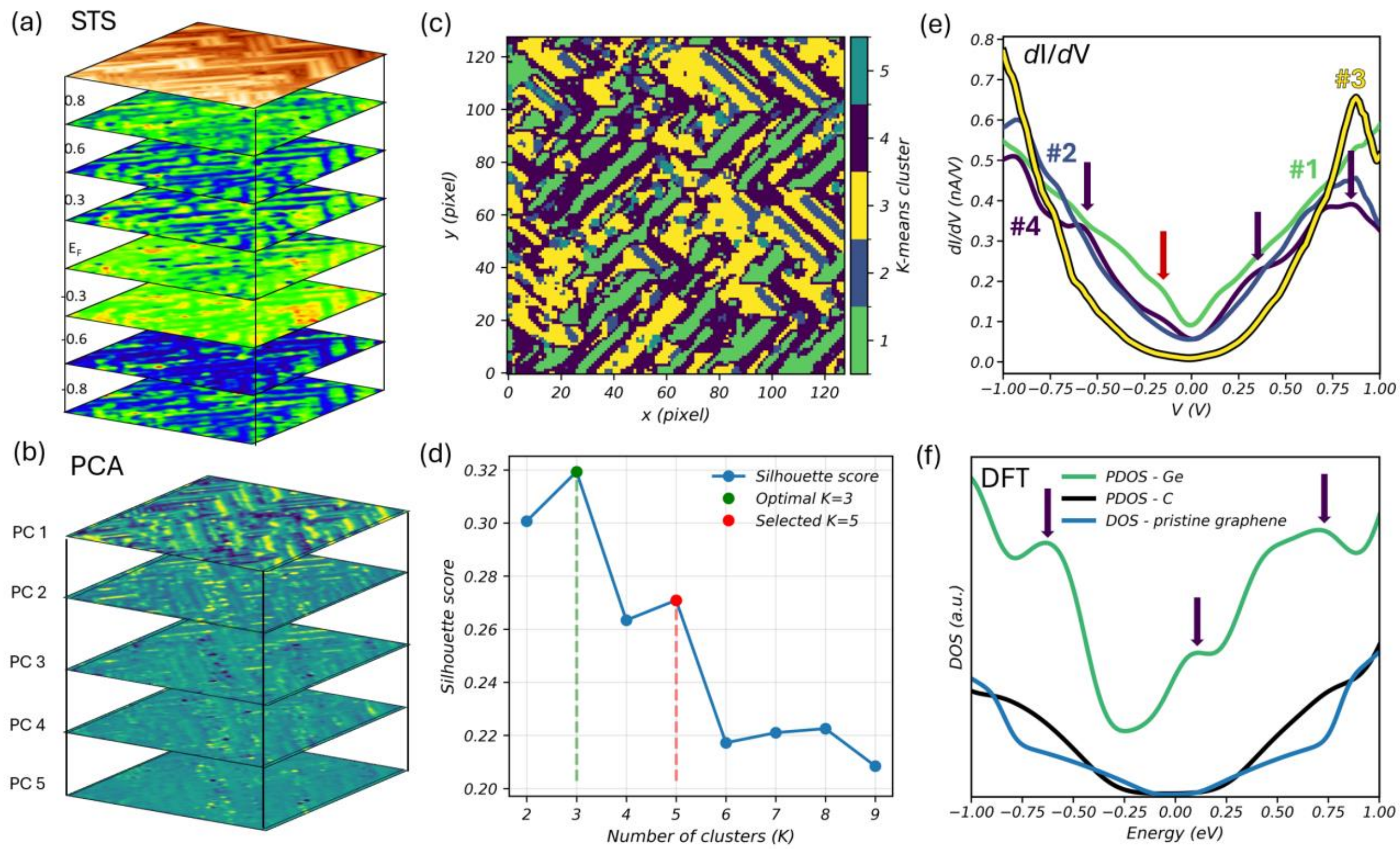

***Figure 4. STS analysis of the graphene/Ge(001) interface***. *(a) Representative differential-conductance (dI/dV) maps selected from a 128-channel hyperspectral STS dataset. (b) PCA results. Reducing the dataset to five principal components retains the main spectral features while minimizing noise. (c) K-means clustering of the PCA-compressed data, identifying five spatial clusters with distinct local spectroscopic responses. (d) Mean silhouette score as a function of the number of clusters, K. (e) Characteristic dI/dV spectra averaged over the selected clusters, illustrating spatial variations in the local density of states (LDOS). The red arrow marks a near-Fermi-level maximum tentatively assigned to a graphene-derived contribution associated with n-type charge transfer. (f) Calculated density of states (DOS) and partial density of states (PDOS) for graphene/Ge(001), together with the DOS of pristine graphene for comparison. The DOS and PDOS curves were scaled for visual clarity to facilitate comparison of their spectral shapes and peak positions with the experimental STS spectra. Navy arrows in panels (e) and (f) indicate features consistent with contributions from Ge-derived states and graphene states modified by interactions with germanium.*

While ARPES and DFT establish the global direction of charge transfer and provide a microscopic description of the pristine interface, they do not resolve how the graphene-Ge coupling varies locally across the nanofaceted surface. To investigate these nanoscale variations, we combined spatially resolved scanning tunneling spectroscopy (STS) with machine-learning-assisted analysis of a three-dimensional spectral dataset composed of 128 energy-resolved images (Figure 4a; see Supporting Information for details – Figure S3 and S4).

Dimensionality reduction was first performed using principal component analysis (PCA)[55], compressing the dataset from 128 channels to five principal components while retaining the essential spectral information (Figure 4b). The reduced data were subsequently classified using the K-means clustering algorithm[56]. The number of clusters was evaluated using the silhouette score.[57] Although lower K values provide a more compact global partitioning of the dataset, K = 5 was deliberately selected because it enables the separation of physically meaningful minority regions that are not resolved at lower K values (see Supporting Information, Figures S5–S7). The resulting classification (Figure 4c) reveals five distinct, non-overlapping spatial regions with different local spectroscopic responses. However, one of these clusters exhibits a mixed spectral response and does not represent an additional distinct electronic regime. Therefore, the subsequent discussion focuses on four characteristic electronic response types. Two of them are consistent with regimes previously identified by AFM/LC-AFM and STS studies: weakly interacting, nearly free-standing graphene on the tops of nanofacets and strongly coupled graphene in the inter-facet regions. The remaining response types correspond to an intermediate coupling regime and to narrow nanoribbon-like regions with distinct electronic features. Representative dI/dV spectra used to assign these local electronic regimes are discussed below.

Representative dI/dV spectra extracted from the selected K-means clusters are shown in Figure 4e. The calculated partial density of states (PDOS) for graphene and Ge (Figure 4f) provides a qualitative reference for interpreting the experimental spectra. Because the theoretical model represents an idealized, flat graphene/Ge(001) interface, direct correspondence with experiment is expected primarily for regions exhibiting stronger graphene–Ge coupling. Experimentally, the spectra reveal a gradual evolution of the graphene–germanium interaction: (i) weakly interacting, nearly free-standing graphene on nanofacet tops (#1, green curve), (ii) an intermediate coupling regime along facet edges (#2, navy curve), (iii) strongly coupled graphene in inter-facet regions (#3, yellow curve), where the graphene electronic structure is substantially modified by the substrate and (iv) nanoribbon-like graphene regions located on a small fraction of narrow facets (#4, blue curve). The fifth cluster (#5, teal) exhibits a mixed spectroscopic response and cannot be assigned to a distinct physical regime and it is therefore treated as an unclassified region in the subsequent discussion. Comparing the intermediate- and strong-coupling spectra with the calculated PDOS/DOS reveals pronounced dI/dV maxima at approximately −0.55, +0.34 and +0.80 eV relative to $E_F$, marked by blue arrows. These features are consistent with contributions from Ge-derived states and graphene states modified by interaction with germanium. The near-Fermi-level maximum at approximately −0.16 eV, marked by the red arrow, is tentatively assigned to a graphene-derived contribution associated with the n-type charge transfer identified by ARPES and DFT.

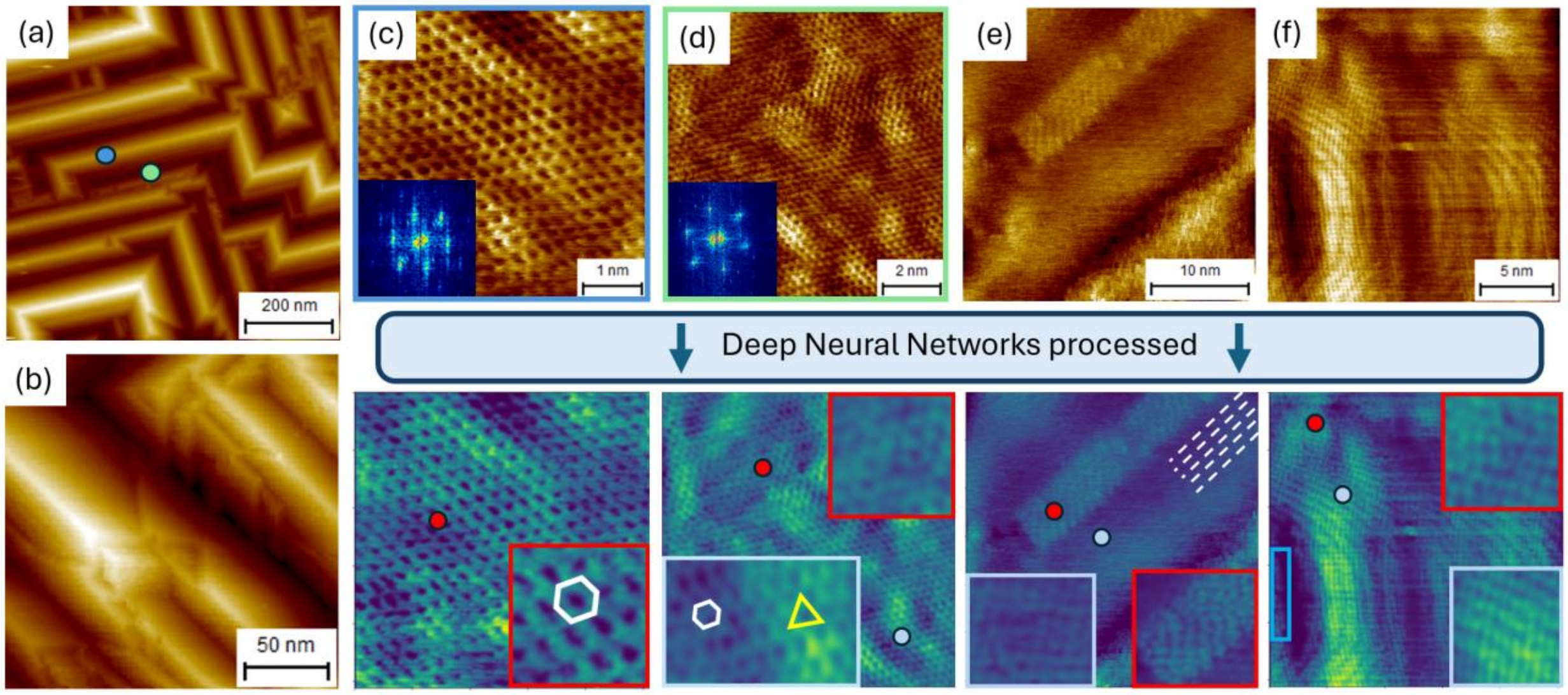

***Figure 5. STM topography of the graphene/Ge(001) surface.*** *(a) Large-scale STM topography showing the faceted surface morphology. (b) High-resolution view of an individual nanofacet. (c) Atomic-resolution image from a facet top (inset: 2D FFT), with the corresponding DNN-processed version (lower panel) highlighting the pristine hexagonal lattice (red dot inset). (d) Structural complexity between adjacent facets. The DNN-processed image (lower panel) reveals a transition from a disordered configuration (upper red inset) to a coexistence of hexagonal graphene and a graphitic-reconstruction (lower gray inset). (e) Nanoribbon-like graphene region on a facet top exhibiting pronounced spatial contrast variation. The insets show atomic-scale details within and adjacent to the ribbon-like region. (f) Magnified view of periodic STM-contrast modulations near the ribbon boundary. The DNN-processed image highlights the transition from an undistorted lattice away from the boundary to a periodically modulated region with a characteristic periodicity of ≈0.5 nm.*

To further clarify the structural origin of the local spectroscopic variations identified by STS and K-means clustering, high-resolution STM imaging was performed (Figure 5). The overall facet morphology is shown in Figure 5a–b. On the tops of the nanofacets (marked by blue dot on Fig. 5a), the STM images reveal well-ordered, nearly free-standing graphene (Figure 5c), consistent with the weakly interacting regions identified by STS and ML analysis. In contrast, the inter-facet areas (marked by the green dot on Fig 5a), exhibit complex topography and electronic heterogeneity as is shown on Figure 5d. To analyze these atomic-scale structures in detail, we employed a deep neural network (DNN) implemented via the AtomAI module[58] within the Pycroscopy[59] framework. The network was trained on a graphene-specific dataset[58] augmented with synthetic sinusoidal noise to mimic the dominant scanning artifacts observed in our experimental STM images. It was subsequently used to denoise atomic-resolution raw STM images while minimizing artifacts associated with conventional processing methods (Raw data are provided in the Supporting Information – Figure S8). The ML-enhanced images clearly distinguish ordered and disordered regions. While consistent with conventionally processed data, the ML approach reveals additional subtle features that are not readily accessible using standard filtering techniques and were therefore used for detailed structural analysis. The upper inset in Figure 5d (ML-processed image)(red rectangle taken from area marked by red dot) shows a region with disrupted atomic periodicity, while the lower inset (gray rectangle taken from area marked by gray dot) displays a mixture of hexagonal graphene-like and graphitic-like for AB layers stacking structures where every second atom is visible[60] These findings directly correlate with the spectroscopic cluster analysis and explain the complex electronic response across the sample.

Interestingly, approximately 10% of the nanofacets host narrow graphene regions with a nanoribbon-like morphology (Figure 5e). In the DNN-processed STM image, the red and gray rectangles mark two spatially distinct contrast regimes, with the gray region exhibiting reduced STM intensity relative to the adjacent region highlighted in red. This contrast indicates that the ribbon-like feature and the surrounding graphene represent distinct local structural and electronic environments rather than a uniformly ordered graphene layer. The white dashed lines indicate the long-axis direction of the ribbon-like region. The occurrence of such structures is consistent with the previously reported anisotropic CVD growth of graphene on Ge(001), where preferentially oriented graphene nanoribbons can form during growth and subsequently merge into a continuous graphene layer.[61] The persistence of locally confined ribbon-like regions within the continuous graphene layer therefore provides a plausible structural origin for the minority spectroscopic response identified by the STS clustering analysis and indicates that their boundaries may constitute distinct local electronic environments.

A closer inspection of the ribbon boundaries (Figure 5f) reveals a pronounced periodic modulation of the STM contrast with a characteristic length scale of approximately 0.5 nm. The modulation is localized near the ribbon boundary and extends several nanometers into the surrounding graphene, whereas regions farther from the boundary retain the regular graphene lattice. The observed localization indicates that the ribbon boundary perturbs the local electronic and/or structural environment. Edge-induced quantum interference provides one possible origin of this modulation. Similar interference patterns associated with intervalley backscattering have previously been observed by STM at graphene edges, including armchair graphene nanoribbons grown directly on Ge(001).[61,62] However, the present data do not allow this mechanism to be distinguished unambiguously from a structural superlattice. In particular, the local lattice distortions visible in the atomic-resolution images and the possible presence of locally stacked graphene layers make a strain-distorted moiré pattern another plausible interpretation. Strain, lattice relaxation and out-of-plane corrugation are known to distort the geometry and periodicity of graphene moiré superlattices and can substantially modify their local electronic structure.[63–65] Thus, the periodic structures observed near the ribbon-like regions are consistent with a locally perturbed electronic state, potentially involving edge-induced interference, a distorted moiré superlattice or a combination of both effects.

Our combined ARPES, DFT, STS, STM and machine-learning analysis reveals that the graphene/Ge(001) interface does not follow a simple dichotomy between weakly and strongly interacting graphene. Instead, it exhibits a continuum of coupling regimes governed by local morphology. These regimes range from nearly free-standing graphene on nanofacet tops to strongly interacting regions in inter-facet areas, with intermediate coupling at facet edges and localized nanoribbon-like structures. This hierarchy of interaction strengths and charge-transfer magnitudes explains the spatial variation of electronic properties observed across the sample and provides a consistent interpretation of the n-type doping signatures identified by ARPES and DFT. Together, results provide a comprehensive, nanoscale-resolved picture of how graphene interacts with germanium surfaces and offer crucial insights for controlling electronic coupling in graphene on semiconductor architectures.

## Conclusions

We investigated the graphene/Ge(001) interface using a correlative framework combining global photoemission, local Scanning probe microscopies, first-principles theory and machine-learning analysis. This approach provides a comprehensive picture of how interfacial structure governs work function, charge transfer and electronic coupling across multiple length scales.

We establish that the work function of graphene/Ge(001) is intrinsically non-uniform. DFT calculations yield a clean-interface value of 4.11 eV, while UPS measurements give a lower global value of ~3.9 eV. This discrepancy arises from electrostatic patch potentials associated with nanoscale variations in graphene–substrate interaction, consistent with KPFM measurements and the broadened UPS cutoff.

A consistent picture of charge transfer mechanism is obtained from both experiment and theory. ARPES and DFT indicate n-type doping, with the Dirac point located below $E_F$, confirming electron transfer from Ge to graphene. In contrast, modeling of an oxidized interface predicts a reversal to p-type doping, demonstrating the sensitivity of charge transfer to interfacial chemistry. The absence of detectable oxidized Ge in our samples supports the assignment of the observed n-type doping to the non-oxidized graphene/Ge interface.

At the nanoscale, machine-learning-assisted classification of the STS data resolves four distinct electronic classes that correlate with the local surface structure: (i) nearly free-standing graphene on the tops of nanofacets, (ii) partially coupled graphene at facet edges, (iii) more strongly coupled graphene in inter-facet regions and (iv) nanoribbon-like regions located on a minority of small, flat facets.

In summary, our study shows that the electronic response of graphene/Ge(001) is controlled by coupled factors: interfacial oxidation and charge transfer, nanofaceted morphology and the associated work-function landscape and local graphene–Ge coupling. That helps reconcile the different doping polarities reported in the literature. The graphene/Ge interface does not behave as a spatially uniform system. Instead, local variations in surface morphology and graphene–substrate coupling give rise to a spatially modulated electronic landscape with distinct local coupling regimes and nanoscale variations in the STM/STS response. These findings demonstrate that the electronic response of graphene/Ge can be tailored through control of the synthesis conditions, interfacial morphology and local coupling strength.

Our results provide practical guidelines for the development of graphene/Ge-based electronic and optoelectronic devices, where spatial variations in conductivity and local electronic coupling may be either minimized or deliberately engineered. They are also relevant for the use of Ge-grown graphene as a transferable protective coating for air-sensitive materials. The feasibility of this approach is further supported by atomically resolved STM measurements of graphene delaminated from Ge(001) and transferred onto $WTe_2$, which show that the characteristic graphene honeycomb lattice is preserved after transfer (Fig. S9). Strong local graphene–substrate interactions may promote defect formation during delamination, whereas synthesis strategies designed to reduce and homogenize the interfacial coupling may facilitate the transfer of higher-quality graphene layers. Controlled growth on Ge(001) therefore offers a scalable route not only toward graphene-based semiconductor architectures but also toward large-area encapsulation platforms for surface-sensitive materials.

## Methods

**Graphene growth on Ge(001)/Si(001)** Graphene films were synthesized by chemical vapor deposition (CVD) in a 6-inch Aixtron Black Magic reactor following established procedures. The substrates consisted of (001)-oriented epitaxial Ge layers grown on Si(001) wafers. Methane ($CH_4$), diluted in Ar/$H_2$ (20:1), was used as the carbon precursor. Prior to growth, Ge substrates were annealed in flowing hydrogen to remove native oxides. Growth and annealing conditions were consistent with previously reported wafer-compatible Ge(001) protocols.[17]

After synthesis, the samples were transferred under inert atmosphere to a Multiprobe P system (Omicron/Scienta Omicron) equipped with a variable-temperature atomic force microscope (VT-AFM), scanning tunneling microscopy (STM), ultraviolet and X-ray sources and a Phoibos 150 hemispherical energy analyzer. Prior to measurements, the samples were annealed at approximately 600 K for 30 min under ultra-high vacuum (UHV, low $10^{-9}$ mbar) to remove physisorbed contaminants.

**Work function and photoemission measurements**

Work function (WF) reference materials: Au/mica (300 nm) and highly oriented pyrolytic graphite (HOPG) were used to verify analyzer calibration. Au was cleaned by sputter–anneal cycles, HOPG was cleaved in UHV immediately before use. Ultraviolet photoelectron spectroscopy (UPS) spectra were acquired using a He I source (21.22 eV). The sample was mounted perpendicular to the analyzer entrance. The secondary-electron cutoff (SEC) was measured with a 9 V bias to enhance low-energy emission. WF values were extracted by linear extrapolation of the SEC, the calibration was conducted at the Fermi level of Au(111). Reference measurements performed under the same experimental conditions yielded a WF of 4.5 eV for the Au reference sample.

**Low-energy electron microscopy (LEEM)**

LEEM measurements were performed under UHV to examine surface morphology and nanoscale WF variations across the graphene/Ge(001) interface, intensity–voltage (I–V) curves and WF maps were collected using standard acquisition protocols.

**Kelvin probe force microscopy (KPFM)**

Local contact potential difference (CPD) maps were obtained using the Multiprobe P (VT-AFM) system operated under UHV and with a separate NT-MDT NTEGRA Aura microscope for ambient measurements.

**UHV KPFM:** Measurements were carried out in frequency-modulation (FM) mode using doped-Si probes (NanoAndMore) controlled by the Matrix system with Nanonis electronics, allowing simultaneous acquisition of topography and CPD with bias applied to the tip during the measurement. CPD maps were obtained simultaneously with topography noncontact-AFM (nc-AFM) images.

**Ambient KPFM:** Measurements were performed in amplitude-modulation (AM) mode. In both cases, only relative CPD contrasts were analyzed, due to calibration uncertainty, absolute WF values were not extracted. CPD line profiles were used to quantify potential differences between nanofacet tops and inter-facet valleys under UHV and ambient conditions.

**Scanning tunneling microscopy (STM) and spectroscopy (STS)**

STM and STS were performed in the same UHV Multiprobe P (VT-AFM) system described above. Electrochemically etched tungsten and mechanically cut PtIr tips were used. STM imaging established facet morphology, atomically resolved images were collected on facet tops and inter-facet regions. STS, tunneling spectra were recorded point-by-point. Each spectrum contained 128 energy points across a bias window from −1 eV to +1 eV. The complete dataset comprised 128 × 128 spatial points, generating a three-dimensional spectral cube for machine-learning analysis.

**Angle-resolved photoemission spectroscopy (ARPES)**

ARPES experiments were conducted at 130 K using a high brightness He I source (hν = 21.22 eV) and a Phoibos 150 analyzer with a 2D detector. The total energy and angular resolutions were 30 meV and 0.3°, respectively. Measurements were performed on graphene/Ge(001)/Si(001) samples after UHV annealing. Dirac-point positions were determined from momentum-distribution and energy-distribution curves.

**Spectroscopic data analysis and machine learning**

STS data were analyzed using unsupervised machine-learning methods. Dimensionality reduction was performed using Principal Component Analysis (PCA), followed by clustering with the K-Means algorithm. STM images were denoised and segmented using a neural-network-based approach implemented in AtomAI. ML-processed data were validated against raw images to ensure preservation of atomic-scale features. Data processing and analysis routines were implemented in Python and are publicly available at: *https://github.com/przybysz-p/stm-ml-analyzer*

**Density functional theory (DFT) Calculations**

First-principles density functional theory (DFT) calculations for the graphene/Ge(001) interface were performed using the Quantum ESPRESSO package.[66,67] Electron–ion interactions were treated using the projector augmented-wave (PAW) method and wavefunctions were expanded in a plane-wave basis with kinetic energy cutoffs of 60 Ry. Exchange–correlation functionals were defined within the generalized gradient approximation (GGA), using the Perdew–Burke–Ernzerhof (PBE) parametrization.[68] For pristine graphene, a plane-wave kinetic-energy cutoff of 80 Ry was used and the Brillouin zone was sampled using an 18 × 18 × 1 k-point mesh. **Supercell and Geometry:** The graphene/Ge(001) supercell was constructed using MedeA, based on the interface model previously reported by Dąbrowski et al.[43]. The clean interface was modeled as a graphene monolayer on Ge(001) with a (2 × 1) dimer reconstruction. The supercell contained 128 atoms (contain 48 carbon atoms, 72 germanium atoms, and 8 hydrogen atoms). The in-plane dimensions of the supercell were [[7.39Å, 0, 0], [0, 17.08Å, 0], [0, 0, 37 Å]] and included a vacuum spacing of ≥ 20 Å. We constructed a commensurate, untwisted graphene/Ge(001) heterostructure by matching a (2 × 4) Ge(001) surface supercell with a (3 × 4) rectangular graphene supercell. The graphene lattice was kept essentially unstrained, while commensuration was achieved by applying an anisotropic in-plane strain of approximately 7% to the Ge substrate. The H atoms were used to passivate dangling bonds at the bottom Ge surface and suppress artificial electronic states associated with the finite slab termination. For oxidized interfaces, oxygen atoms were placed on surface Ge dimers following established Ge–O bonding configurations[45,54]. The upper Ge layers and adsorbates were fully relaxed in both configurations, until residual forces were below $10^{-4}$ Ry/Bohr, including van der Waals interactions treated with the DFT-D3 approach[69]. The Brillouin zone was sampled using a 4 × 4 × 1 k-point mesh for heterostructure. Band unfolding was carried out using the BandUP(py) code[70–72].

**Supporting Information**

Additional UPS, LEEM, XPS, and LEED characterization; normalized and non-normalized STS maps; PCA and K-means clustering analysis; original STM topographies used for DNN processing; and atomic-resolution STM characterization of graphene transferred onto $WTe_2$.

**Acknowledgments**

This work was financially supported by the National Science Centre, Poland under projects 2018/30/E/ST5/00667 (P.D. and P.P), 2025/58/E/ST11/00213 (I.P. and J.S.), 2025/59/D/ST11/02390 (I.L) and 2020/38/E/ST3/00293 (M.R.). P.P. acknowledges support from the "Smarter, Faster, Better! – Internationalisation of Doctoral Schools at the University of Łódź" project funded by STER NAWA - Internationalisation of Doctoral Schools Programme. J.S. acknowledges the Rosalind Franklin Fellowship from the University of Groningen. The calculations were carried out on the Dutch national e-infrastructure with the support of SURF Cooperative (EINF-8924) and on the Hábrók high-performance computing cluster of the University of Groningen. We gratefully acknowledge Bartłomiej Dąbrowski, Aleksander Karliński and Jakub Bogołębski for their assistance with the preliminary development and testing of the STS data-analysis routines.

**Supplementary Materials: Electronic Coupling and Charge-Transfer Landscape of Graphene on Ge(001)/Si(001): Multiscale Analysis Assisted by Machine Learning**

## Work-Function, XPS, and LEED Characterization

Since tabulated WF values, even for well-characterized materials, can vary significantly due to impurities, surface reconstructions, and local environmental conditions, it is strongly recommended to measure reference samples using global techniques such as UPS rather than relying solely on literature values. In our study (Table 1), we present UPS-derived WF values for our reference samples, Au(111) and HOPG, as well as for graphene/germanium, and graphene on 4H-SiC. When a well-defined Fermi edge was not observed, the Fermi-level position determined from the Au reference was used for energy calibration.

To additionally validate the estimated local WF variations in graphene/Ge(001), LEEM intensity–voltage curves were recorded across the transition region between mirror electron microscopy (MEM) and LEEM. The onset shift observed between representative regions of the nanofaceted surface allows the local WF difference between nanofacet tops and inter-facet regions to be estimated at approximately 0.2 eV. Although the LEEM-derived estimate is larger than the CPD contrast measured by KPFM, both measurements indicate spatial variations in the local work function across the nanofaceted surface. The LEEM image inset in Figure S1 shows the corresponding contrast between regions with different local WFs, supporting the interpretation of the graphene/Ge(001) surface as laterally heterogeneous in electrostatic potential.

Table 1: *Comparison of UPS-measured work function (WF) values for Au(111), HOPG, graphene/SiC(4H) and graphene/germanium samples.*

| | Au(111) | HOPG | G/4H-SiC | G/Ge(001) |
|---|---|---|---|---|
| WF [eV] | 4.55 | 4.5 | 4.45 | 3.9 |

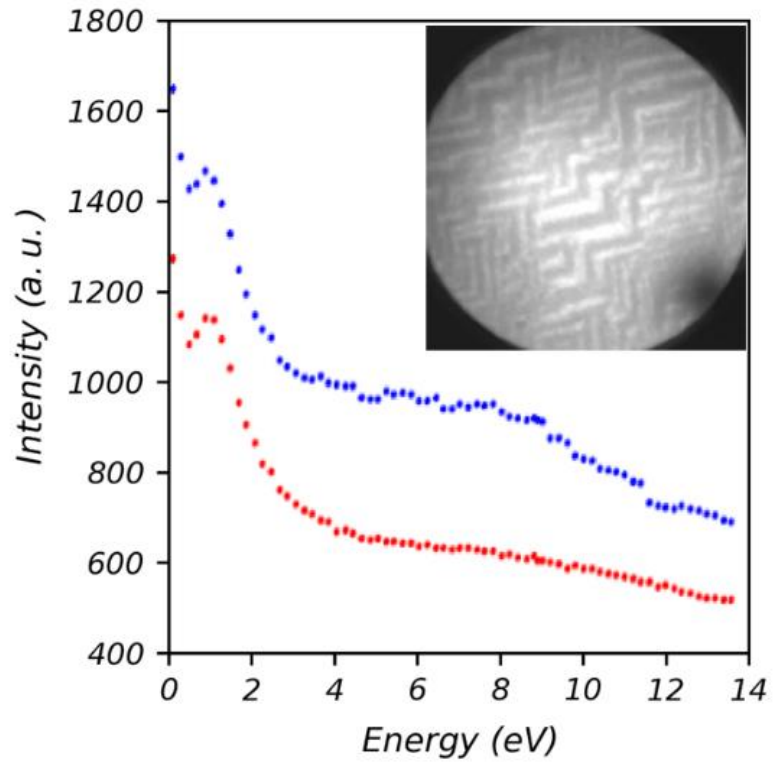

***Fig.S1.*** *LEEM intensity curves and an image indicating WF inhomogeneities across the sample due to nanofacet formation.*

The global XPS measurements shown in Fig. S2 confirm the presence of the graphene C 1s core-level peak and the Ge 2p signal from the germanium substrate. No evidence of substrate oxidation is observed in these spectra. In addition, the LEED pattern recorded for graphene/Ge(001) reveals two predominant graphene domain orientations, indicated by the black and green hexagons and arrows. Diffraction spots originating from the Ge substrate are marked by red circles. The blue dotted circles indicate different orientations of the Ge surface with respect to the LEED spectrometer, arising from the nanofaceted morphology of the substrate.

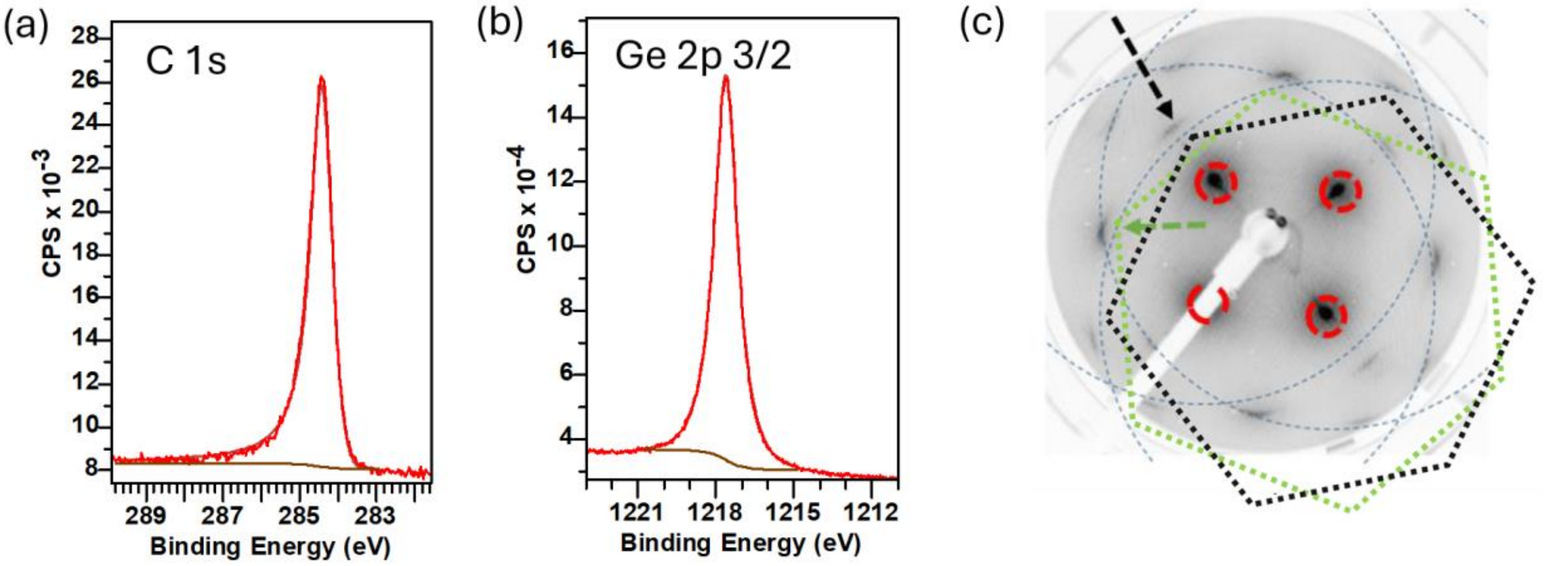

***Fig. S2.*** *(a) XPS spectrum showing the C 1s peak, confirming the presence of graphene. (b) XPS spectrum showing the Ge peak from the germanium substrate. (c) LEED pattern confirming the presence of Ge(001) and graphene, with different domain orientations attributed to nanofacet formation.*

**STS**

To probe the local density of states, STS measurements were performed. After numerical differentiation, normalized differential conductance maps, $(dI/dV)/(I/V)$, were obtained. The original data used to prepare Fig. 4 in the main manuscript are shown in Fig. S3. For comparison, we also present the data without Feenstra normalization, which exhibit enhanced contrast due to slight convolution with sample height variations (Fig. S4).

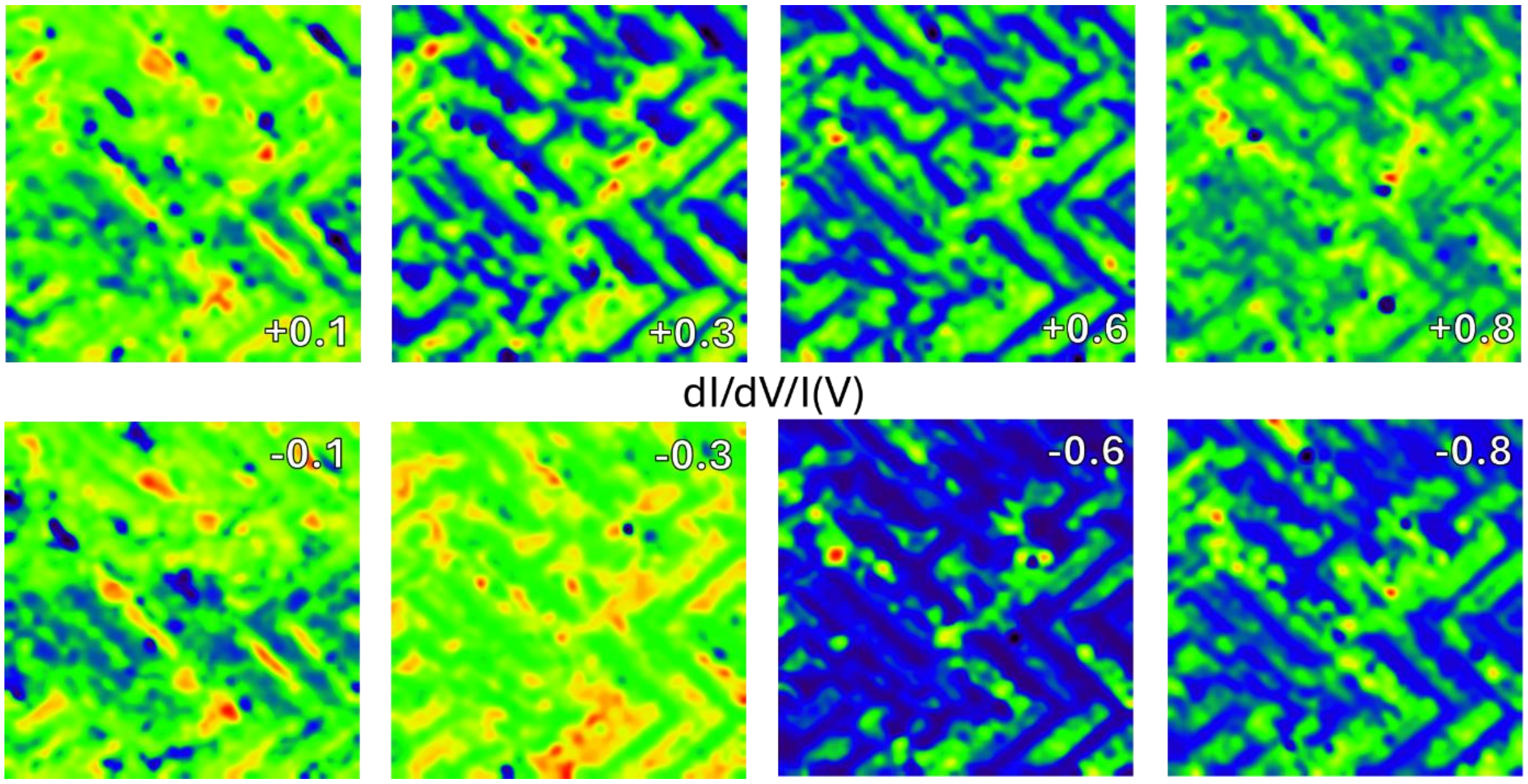

***Fig. S3***. *Scanning tunneling spectroscopy (STS) maps of graphene/Ge(001) obtained after numerical differentiation and Feenstra normalization, highlighting spectroscopic differences between nanofacet tops and inter-facet valleys. These data were used to prepare Fig. 4(a) in the main text.*

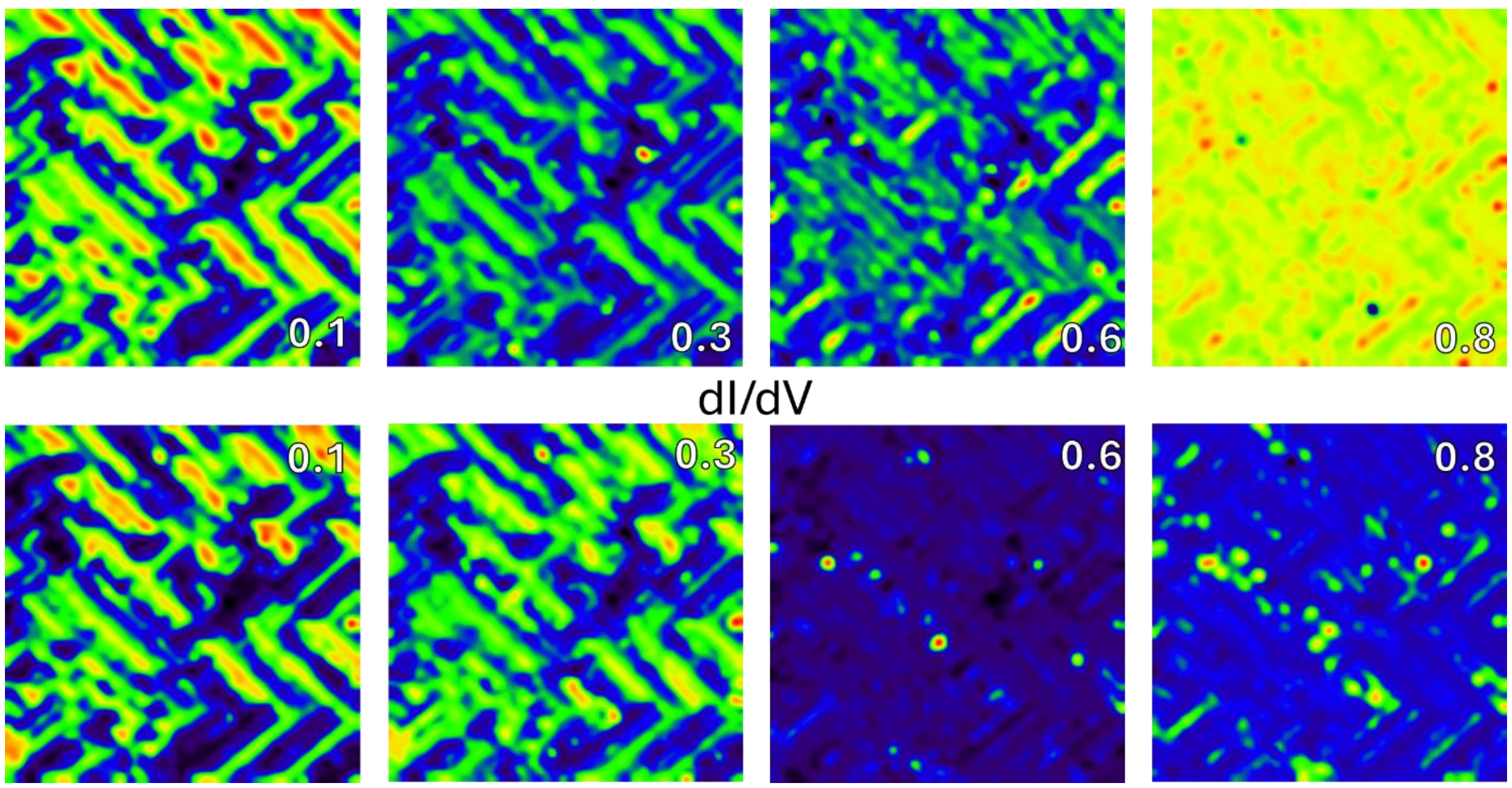

***Fig. S4***. *Scanning tunneling spectroscopy (STS) maps of graphene/Ge(001) obtained after numerical differentiation, without Feenstra normalization.*

**PCA and K-means clustering**

For the normalized data shown in Fig. S3, PCA-based dimensionality reduction was also performed. Figure S5 presents the five principal-component score maps used in the analysis presented in Fig. 4 of the main text.

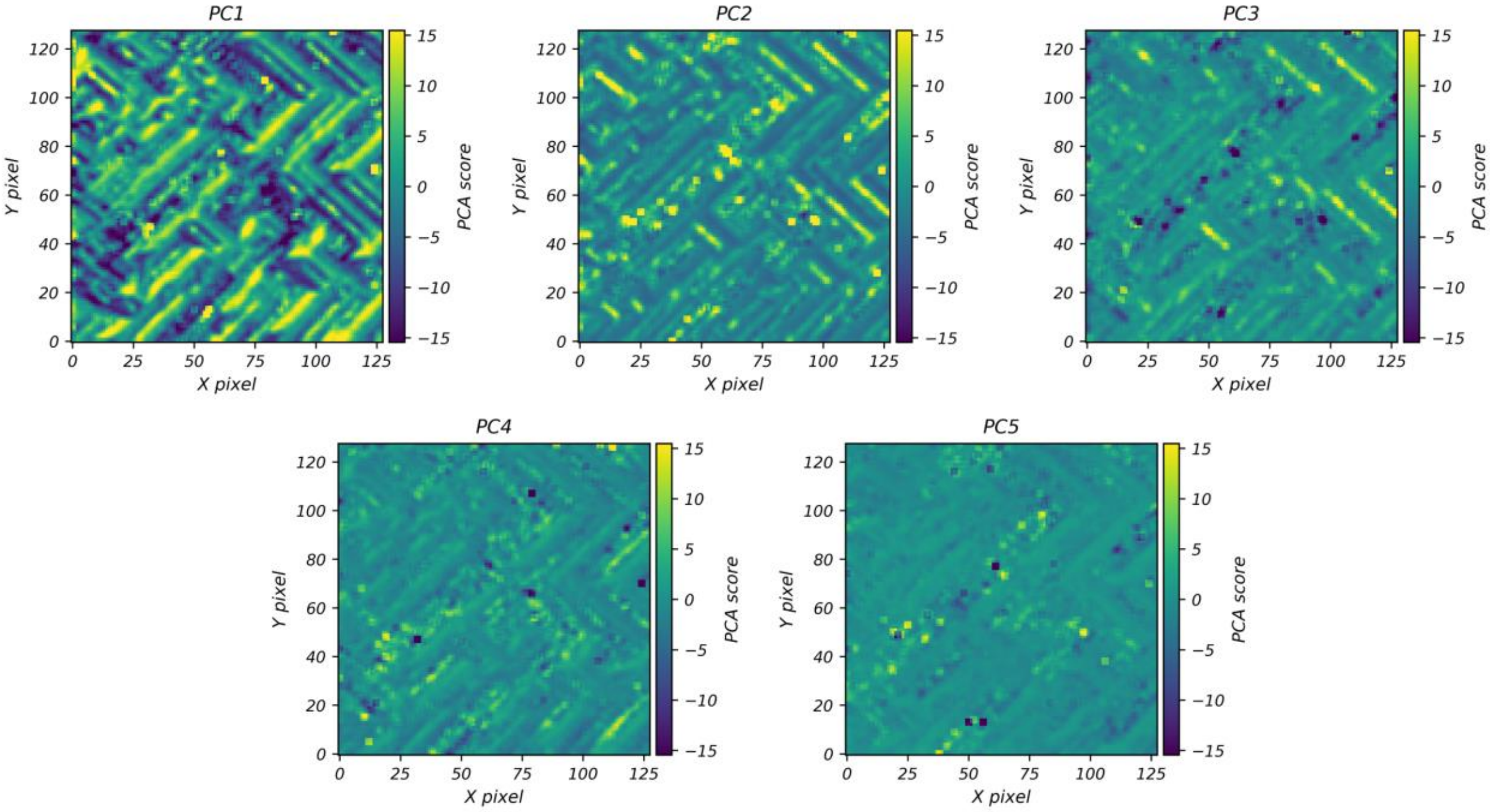

***Fig.S5.*** *Results of principal component analysis (PCA) used to reduce the normalized differential conductance dataset to five principal components while retaining all key spectral features. Those data were then classified by K-Means.*

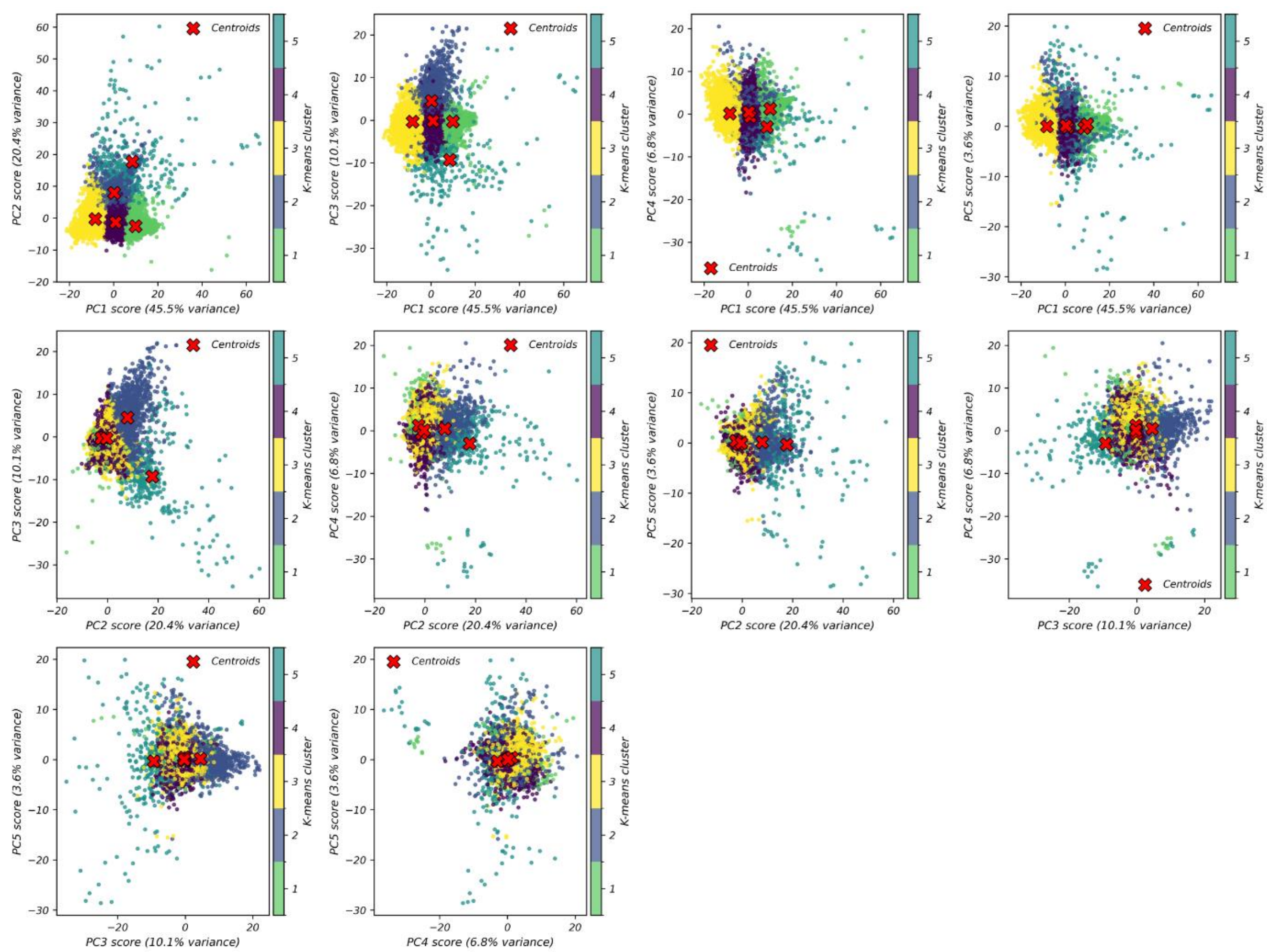

***Fig.S6.*** *Pairwise projections of the STS spectra onto the first five principal-component score axes, colored according to the K = 5 K-means cluster labels.*

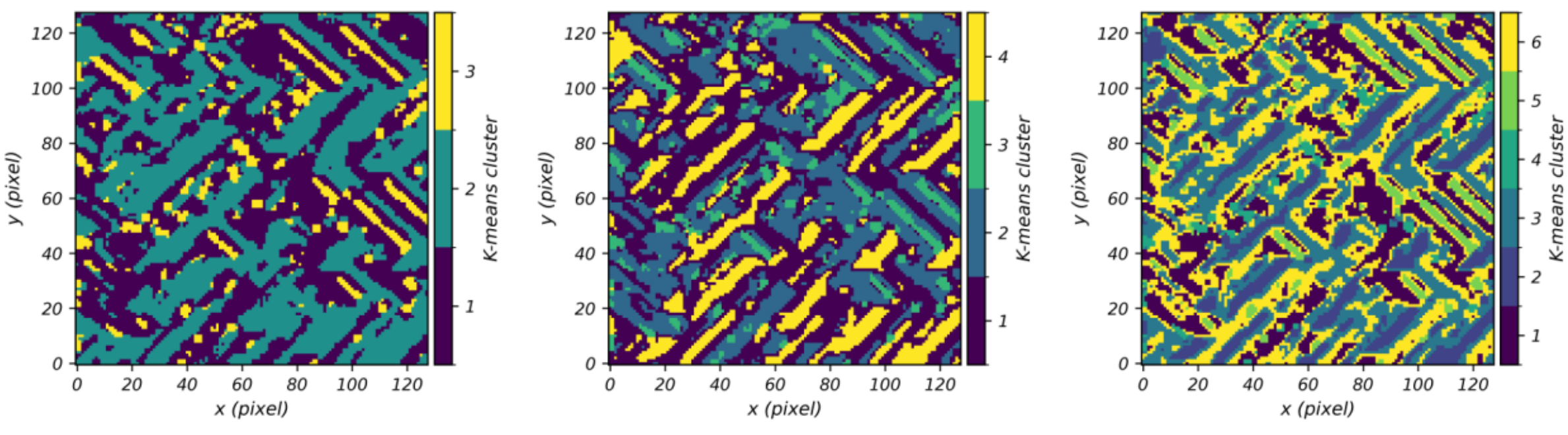

***Fig.S7.*** *Spatial K-means clustering maps for K = 3, 4, and 6.*

The K = 5 partition was selected as the best compromise between spatial resolution and physical interpretability. K = 3 and K = 4 provide an overly coarse segmentation and do not resolve minority regions, whereas K = 6 leads to over-segmentation without a clearer physical assignment. Although one of the five clusters exhibits a mixed spectral response, the remaining clusters separate the main local electronic regimes discussed in the manuscript. K-means clustering was performed independently for each K value. As a result, cluster labels and colors for K = 3, 4, 5, and 6 should not be interpreted as representing the same spatial or electronic regions. The physical meaning of the clusters must be assessed separately for each partition.

## STM

During STM data analysis, several processing steps were applied, including plane correction, averaging, FFT filtering, and scratch removal, in order to improve data visibility. In parallel, the same data were processed using a neural network implemented in AtomAI. For this analysis, only plane correction was applied to the STM topography images prior to the DNN-based processing. Figure S8 shows the STM topography after plane correction only, i.e., in a form close to the raw experimental data.

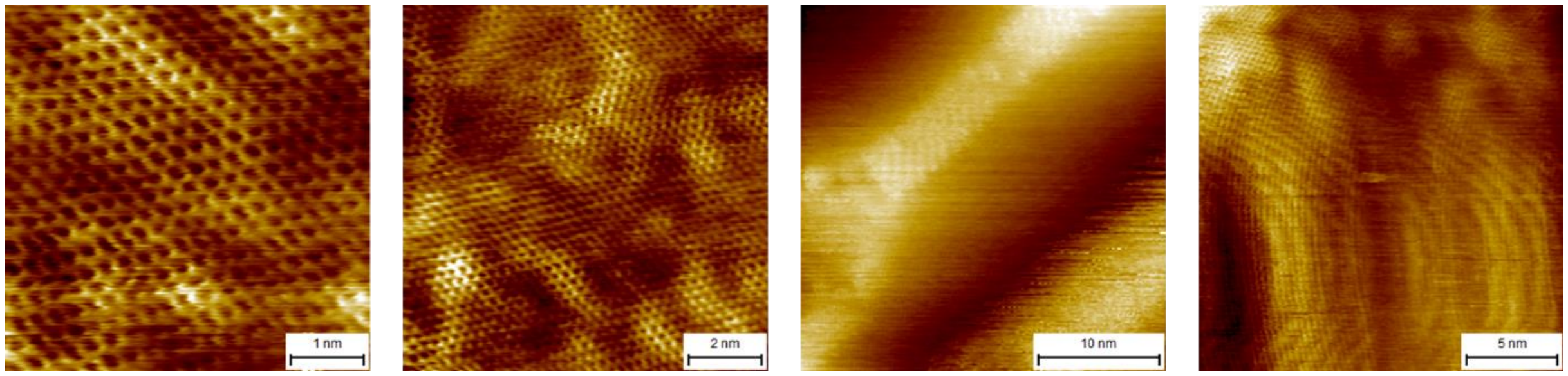

***Fig.S8.*** *Original STM topographies used for subsequent analysis. Only plane correction was applied.*

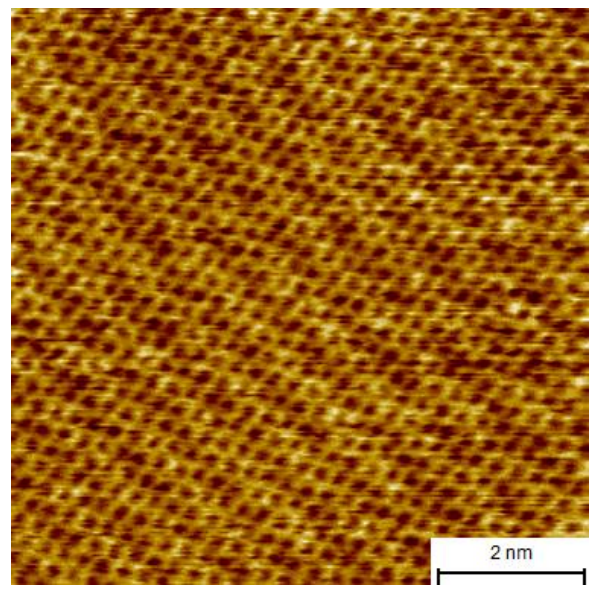

***Fig. S9.*** *Atomically resolved STM topography of graphene delaminated from Ge(001) and transferred onto a* $WTe_2$ *substrate, showing a well-resolved honeycomb lattice.*